\documentclass[prb]{revtex4}

\usepackage{graphicx}
\usepackage{epsfig}
\usepackage{dcolumn}
\usepackage{bm}
\usepackage{amsfonts}
\usepackage{amsmath}
\usepackage{textcomp}
\usepackage{mathtools, bm}
\usepackage{amssymb, bm}

\usepackage[a4paper,includeheadfoot,margin = 2.2cm]{geometry}	

\usepackage{appendix}

\usepackage{tikz}
\usetikzlibrary{arrows}
\usetikzlibrary{decorations.markings}
\tikzstyle directed=[postaction={decorate,decoration={markings,
    mark=at position .65 with {\arrow{latex}}}}]

\usepackage{tikzSettings}

\usepackage{hyperref}

\hypersetup{urlcolor=blue, colorlinks=true}  

\hypersetup{colorlinks=true,linkcolor=blue}

\begin{document}

\title{Effective resistance of random percolating networks of stick nanowires : functional dependence on elementary physical parameters}
\author{Robert Benda$^{1,2}$} \email{robert.benda@polytechnique.org}, \author{B\'ereng\`ere Lebental$^{1,3}$}, \author{Eric Canc\` es$^{2}$ }
 \affiliation{$^1$ LPICM, CNRS, Ecole Polytechnique, 91128 Palaiseau, France}
\affiliation{$^2$ CERMICS, Ecole des Ponts and INRIA, Universit\'e Paris-Est, 6-8 avenue Blaise Pascal 77455 Marne-la-Vall\' ee, France}
\affiliation{$^3$ Universit\'e Paris-Est, IFSTTAR, 14-20, Boulevard Newton, 77420 Champs-sur-Marne, France}

\date{\today}

\begin{abstract}
We study by means of Monte-Carlo numerical simulations the resistance of two-dimensional random percolating networks of stick, widthless nanowires. We use the multi-nodal representation (MNR) \cite{Rocha2015} to model a nanowire network as a graph. 
We derive numerically from this model the expression of the total resistance as a function of all meaningful parameters, geometrical and physical, over a wide range of variation for each. We justify our choice of non-dimensional variables applying Buckingham $\pi-$theorem. The effective resistance of 2D random percolating networks of nanowires is found to write as $R^{eq}(\rho,R_c,R_{m,w})=A\left(N,\frac{L}{l^{*}}\right) \rho l^*  + B\left(N,\frac{L}{l^{*}}\right) R_c+C\left(N,\frac{L}{l^*} \right) R_{m,w}$ where $N$, $\frac{L}{l^{*}}$ are the geometrical parameters (number of wires, aspect ratio of electrode separation over wire length) and $\rho$, $R_c$, $R_{m,w}$ are the physical parameters (nanowire linear resistance per unit length, nanowire/nanowire contact resistance, metallic electrode/nanowire contact resistance). The dependence of the resistance on the geometry of the network, one the one hand, and on the physical parameters (values of the resistances), on the other hand, is thus clearly separated thanks to this expression, much simpler than the previously reported analytical expressions.

\end{abstract}

\maketitle 

\vspace{2.5cm}

\section{Introduction}

The study of random percolating networks of conducting nanowires has become in the last years a hot topic of investigation. Numerous applications are possible thanks to the outstanding electrical and optical performances of such thin films, combined to their low cost and ease of fabrication. Applications are as diverse as thin film transistors based on carbon nanotubes networks (CNNs) \cite{Li2007,Li2008}, CNNs-based field-effect transistors \cite{Li2009}, transparent conductive (high transmittance, low resistance) thin-film electrodes based on silver nanowires \cite{Rocha2015,OCallaghan2016} or carbon nanotubes (CNTs) \cite{Hecht2006} -- useful for instance in the context of solar cells, the conducting nanowire network acting as a charge carrier collector \cite{Kumar2017a} -- and sensors \cite{Li2009} .
The conductivity, as well as the sheet transmittance of such thin film nanowire networks has been extensively studied experimentally and theoretically \cite{Hecht2006,Nirmalraj2009,Forro2018}.  
These systems are well approximated by random 2D networks, and we study in this paper the dependence of their effective resistance on all structural and physical parameters. We focus on the functional dependence of the resistance on physical parameters, motivated by sensor applications of nanowire networks. In this context, physical parameters, such as nanowire linear resistance, or nanowire/nanowire contact resistance, can vary, due to environment changes. The geometry and structure of the network is thus not sufficient to understand properly the resistance variations. \\

Many parameters impact the total effective resistance of nanowire random networks. Structural parameters such as the density (or coverage), the aspect ratio of electrode separation to wire length, and the alignment of the wires (\textit{i.e.} the statistical distribution of their angles with respect to a fixed direction) are, among others, known to play on the transport properties of the networks, as confirmed experimentally and numerically in \cite{Kocabas2007}.
Simoneau \textit{and al.} studied numerically the influence on the estimated effective resistance of the statistical distribution chosen for angles, lengths, diameters of nanowires (in that case CNTs) \cite{Simoneau2013}, waviness, degree of penetration of CNTs \cite{Simoneau2015}, under the junction dominated assumption (JDA) -- \textit{i.e.} assuming ballistic transport in individual CNTs. The mean tube length, the porosity (or equivalently, the volume fraction of CNTs) and the dimension of the network (distance between the electrodes) are also crucial geometrical parameters well studied in the literature. For instance, Lyons \textit{and al.} \cite{Lyons2008} used an empirical relation relating the mean tube length and the porosity of the network to the total number of junctions, which is shown to mainly drive the film conductivity. The aspect ratio  of the rods themselves (\textit{i.e.} length over diameter), when they are not modeled as widthless sticks, is also a possible varying parameter, whose influence on the total resistance has been studied by Mutiso \textit{and al.} \cite{Mutiso2013}.\\

Concerning physical parameters, the influence of junction resistance (nanowire/nanowire contact resistance, denoted as $R_c$ in the following) on the overall resistance was reported qualitatively, from experimental measurements \cite{Hecht2006}, while a linear relationship between the conductivity and the number of junctions was established in \cite{Lyons2008} for CNTs networks. Several studies addressed the question of the influence of a change of this contact resistance between two nanowires on the overall resistance. Nimalraj \textit{and al.} measured the distribution of junction resistances as a function of the diameter and treatment (acid treatment, annealing, pristine CNTs), and the influence of its variation on the total resistance of the network \cite{Nirmalraj2009}. Finally, Behnam \textit{and al.} \cite{Behnam2007} studied numerically the dependence of the scaling (with respect to the width $L$ of the conducting channel) of the resistance of 2D nanowire networks, on nanotube length, nanotube alignment, and on the ratio of junction to linear resistance (still assuming ballistic transport in CNTs). The linear resistance of nanowires was fixed to the ballistic resistance value, so that its contribution to the whole resistance was not studied in the diffusive regime in Ref. \cite{Behnam2007}, nor did the authors derive a functional dependence of the effective resistance on all the parameters.\\

In the articles previously mentioned, the electronic transport in the percolating network is assumed dominated by the junction resistances, and the contribution of linear resistances, \textit{i.e.} of the wires themselves, are neglected -- Junction Dominated Assumption (JDA). Zezelj \textit{and al.} performed in 2012 a pioneering study of the dependence of conductivity local critical exponents on the junction to linear resistance ratio \cite{Zezelj2012}. da Rocha \textit{and al.} introduced in 2015 the Multi Nodal Representation (MNR) method \cite{Rocha2015}, generalizing the JDA -- mostly used in the previous studies, \textit{e.g} in \cite{Mutiso2013} --, to include wire resistances. In MNR, an electrical circuit made up of randomly dispersed nanowires is represented as a graph, with nodes at the crossing points of nanowires and edges (modelling connecting resistors) in-between. It is also the spirit of the method of nodal potentials already used in Ref. \cite{Wu2004,Kagan2015}, whose idea is to represent each point where the potential is well defined as a node. To account for the resistances across nanowires junctions, each node is then duplicated, yielding two nodes each representing the contact point on one nanowire \cite{Rocha2015,Kim2018} -- every node is thus attached to a given nanowire. A new edge with junction conductance is created in-between. A typical example of construction of the graph is shown Figure \ref{cas_contact_3_voisins_graphe_avec_Rc}.\\ 
da Rocha \textit{and al.} applied this method to derive the contribution of both junction and linear resistances to the total resistance \cite{Rocha2015}. This allows to compute the minimal reachable network resistance. Indeed, for silver nanowires networks, the goal is often to decrease as much as possible the contribution of junction resistances. In the limit of zero junction resistances, the effective resistance is given by the contribution of linear resistances only. In 2016, O'Callaghan \textit{and al.} introduced an equivalence between a random disordered network of nanowires and a regular (square) ordered network, the edges of the associated graph all having the same resistance \cite{OCallaghan2016}. They used effective medium theory arguments, introduced by Kirkpatrick \cite{Kirkpatrick1973}, valid over a certain regime (high above percolation) and exact only in the case of an infinite medium. The authors thus derived a closed-form expression for the homogeneous edge resistance of a square lattice equivalent to (\textit{i.e.} having same total resistance as) the initial disordered network. Incidentally, as they used the MNR method to describe the network, this yielded an expression of the effective resistance as a function of geometrical parameters (number of wires) and physical parameters (with both linear and junction resistances), reported in Table \ref{expressions_analytiques_R_eq_references}. The contact resistance between the metallic electrodes and the nanowires was not included in the model. In 2018,  Kim \textit{and al.} used the same MNR method and a block-matrix representation of Kirchhoff's laws to derive the effective conductivity of 2D random percolating networks of widthless rods \cite{Kim2018}, with three varying geometrical and structural parameters (aspect ratio $\frac{L}{l^{*}}$, length of the rods $l^*$, density of tubes $n$). However, as underlined by the authors, the internal resistance $r_i$ (linear resistance of the wires), although included in the model, was taken constant over all edges of the underlying graph -- while it depends actually on the length of the edge in the diffusive regime. Moreover, the ratio of linear over junction resistance $\frac{r_i}{r_j}$ was fixed for the whole study. Thus, the functional dependence on the physical parameters (linear resistance per unit length of a nanowire, denoted as $\rho$ in the following, junction resistance $R_c$, and metallic electrode/nanowire resistance, denoted as $R_{m,w}$) was not investigated.\\
Finally, Forr\'o \textit{and al.} \cite{Forro2018} also recently derived an analytical expression of the sheet resistance of random 2D networks of nanowires, valid at rather high densities (well above the percolation threshold), as a function of both geometrical parameters (wire density, wire length and electrode separation) and physical parameters $R_c$ (contact resistance) and $R_w$ -- wire resistance, equal in our notations to $\rho l^{*}$ where $l^{*}$ is the length of a wire -- as seen in Eq. (6) of Ref. \cite{Forro2018}, reported Table \ref{expressions_analytiques_R_eq_references}. The authors assumed that the transport properties of the random network was the same as in an equivalent system of individual decoupled wires, with a linear average potential background. For the first time, average individual contact and linear resistance values $R_c$ and $R_w$ were extracted by fitting experimental values of resistances of SWCNTs thin films, of varying geometrical parameters, with $R_c$ and $R_w$ among the fitting parameters. This is worth mentioning as, except thanks to AFM measurements, the individual junction resistances and the average linear resistance cannot be deduced from the measured total resistance of a network. Still, the analytical expression of the total resistance derived by Forro \textit{and al.} does not allow to separate the contributions of all linear resistances and of all the contact resistances to the total resistance (\textit{e.g.} as a sum), contrary to the work of da Rocha \textit{and al.} \cite{Rocha2015}. Moreover, electrode/nanowire resistance $R_{m,w}$ was not treated as an independent variable in this study \cite{Forro2018}, but taken proportional to the nanowire/nanowire contact resistance $R_c$.\\

\begin{table}[!h]\centering
\begin{tabular}{|>{\centering}m{2cm}|>{\centering}m{3.5cm}|>{\centering}m{4.2cm}|>{\centering}m{3.2cm}|>{\centering}m{3.5cm}|}
\hline
 Reference & Variables considered & Formula & Method & Validity  of the expression and drawbacks \tabularnewline
 \hline
 Forr\'o \textit{and al.} \cite{Forro2018} & $R_w=\rho l^*$, $R_c$, $D=\frac{N}{\left( \frac{L}{l^*} \right)^2}$, $n_a \varpropto D$, $r_m$ (see expression in \cite{Forro2018}) & $\hat{R^{eq}}(x=\frac{\rho l^*}{R_c})= \frac{\frac{x}{D}}{\frac{r_m}{2}-\sqrt{\frac{r_m}{2 n_a x}}\tanh\left( \sqrt{\frac{n_a r_m}{2} x} \right) }$ & Equivalent system of decoupled wires with linear average potential background & High density $\frac{N}{N_{cr}} \geq 3$  $\left( N_{cr} \approx 5.64 \left( \frac{L}{l^*} \right)^2 \right)$
 \tabularnewline
 \hline
O'Callaghan \textit{and al.} \cite{OCallaghan2016} &  $g_i=\frac{1}{\rho l_{m}}$, $g_j=\frac{1}{R_c}$, $N$, average segment length $l_{m}$, wire length $l^*$, device size $L$ &  $R^{eq}=R^{*} \frac{N_x}{N_y}$, the expressions of $R^{*}$, $N_x$ and $N_y$ are given below $(*)^{1,2,3}$ &  Effective medium theory  \cite{Kirkpatrick1973} & High above percolation, infinite medium. Assumption of small-world network behavior. \tabularnewline
\hline
Kim \textit{and al.} \cite{Kim2018} & Aspect ratio $\frac{L}{l^*}$, length of the rods $l^*$, density $n$ & No analytical expression & Block-matrix representation of Kirchoff's laws & Ballistic internal wire resistance $r_i$, fixed ratio $\frac{r_i}{r_j}$ (linear to junction resistance) \tabularnewline
\hline
Zezelj \textit{and al.} \cite{Zezelj2012} &  Stick conductance  $G_s = \frac{1}{\rho l^*}$, junction conductance $G_j=\frac{1}{R_c}$, density $n=\frac{N}{\left( \frac{L}{l^*} \right)^2}$, aspect ratio $\frac{L}{l^*}$ &  $R^{eq}=\frac{1}{\sigma} \varpropto b \frac{1}{n} \left(\rho  l^* \right) +\frac{1}{n^2} R_c$ &  Empirical expression (understanding and validation in terms of critical exponents) & Expression of this form in $\rho$ and $R_c$ on the whole density range (with modified geometrical coefficients to include finite-size effects on some ranges). \tabularnewline
\hline
\end{tabular}
\caption{Some "closed-form" expressions derived in the literature (either with effective medium approaches, or empirically with numerical validation) for the resistance of 2D random percolating networks. The parameters $L$, $l^*$, $N$ always refer to the electrode separation, the wire length, and the number of wires in our notation. Devices are assumed of square dimensions $L \times L$. The non-dimensional parameter $x$ is equal to $\frac{\rho l^*}{R_c}$.   
\newline $(*)^1$ $\frac{1}{R^{*}}= \frac{1}{2\left( 3 \alpha N^2 +N \right)}\left[ \alpha N^2 \left( \frac{1}{x}-1 \right)-N \left( \frac{3}{x}+1\right) +\sqrt{\frac{ 12 \left(\alpha^2 N^2-N \right) \left(3 \alpha^2 N^2+N \right) }{x}+\left( \frac{\left(\alpha^2 N^2-3N \right)}{x}-\left( \alpha^2 N^2+N \right)  \right)^2 } \right]$ \newline 
$(*)^2$ $N_x=\frac{L}{l^*}\frac{C}{P_i} \log(6 n_j l^* P_i)$ where $P_i=\frac{2 \alpha N^2-N}{3 \alpha N^2+N}$ and $\alpha
= \frac{0.2027 \pi}{2} (l^*)^2$, $n_j= \alpha N^2$, $C \approx 1$ a constant 
\newline $(*)^3$ $N_y= \frac{ l^* L N }{ \pi}$}
\label{expressions_analytiques_R_eq_references}
\end{table}

The main recent results concerning the functional dependence of the effective resistance of 2D random nanowire networks on physical (and geometrical) parameters -- \textit{i.e.} individual resistances of the components -- are summarized Table \ref{expressions_analytiques_R_eq_references}. We notice that very different expressions of the effective resistance as a function of the physical parameters have been reported, depending on the particular approach used. Moreover, the contact resistance between metallic electrodes and nanowires has never been included in these models.\\

We propose in this paper a different approach to derive an approximate analytical expression of the effective resistance of a random percolating network of nanowires, neither based on percolation nor on effective medium approaches. We study the dependence on all relevant geometrical and physical parameters, for a given random network generating procedure. We also use the MNR method to describe the percolating network of nanowires, including both linear and contact resistances ($\rho$ and $R_c$) as in Ref. \cite{Rocha2015,OCallaghan2016,Kim2018}, and adding the electrode/nanowire contact resistance in the model, for the first time. A closed form expression for the effective resistance, as a function of these physical parameters, is estimated numerically, by means of Monte-Carlo simulations and fitting of the obtained data points. The effective resistance is found to write as simply as $R^{eq}(\rho,R_c,R_{m,w})=A\left(N,\frac{L}{l^{*}}\right) \rho l^*  + B\left(N,\frac{L}{l^{*}}\right) R_c+C\left(N,\frac{L}{l^*} \right) R_{m,w}$, validating numerically (in the particular case of $R_{m,w}=0$) the empirical functional dependence on the linear and junction resistances proposed by Zezelj \textit{and al.} \cite{Zezelj2012}.

\section{Method}

\begin{figure}[!h]
   \begin{minipage}[c]{0.40\linewidth}
   \centering
   \includegraphics[width=6cm,height=5cm]{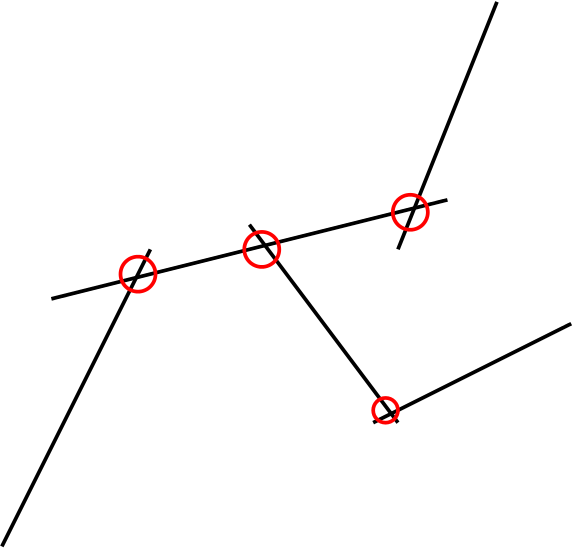}     
  \end{minipage}
  \hfill
 \begin{minipage}[c]{0.55\linewidth}
\begin{center}
\begin{tikzpicture}
\def\l{1.5}
\def\h{1}
\node[sta] (x) at (0*\l,0*\h) {$i_k^1$};
\node[sta] (y) at (0.75*\l,0.75*\h) {$i_k^2$};
\node[sta] (z) at (1.9*\l,2*\h) {$i_h^1$};
\node[sta] (t) at (2.7*\l,1.3*\h) {$i_h^2$};
\node[sta] (w) at (3.8*\l,0.3*\h) {$i_m^1$};
\node[sta] (r) at (4.7*\l,0*\h) {$i_m^2$};
\node[sta] (u) at (3*\l,3.5*\h) {$i_j^1$};
\node[sta] (v) at (3.8*\l,4.2*\h) {$i_j^2$};
\draw (x) -- (y) node [midway,above,color=blue] {$R_c$};
\draw (y) -- (z) node [midway,above,font=\footnotesize,color=green] {$\rho d_{hk}$} ;
\draw (z) -- (t) node [midway,below,color=blue] {$R_c$};
\draw (t) -- (w) node [midway,above,font=\footnotesize,color=green] {$\rho d_{hm}$} ;
\draw (w) -- (r) node [midway,above,color=blue] {$R_c$};
\draw (z) -- (u) node [midway,above,font=\footnotesize,color=green] {$\rho d_{hj}$};
\draw (u) -- (v) node [midway,below,color=blue] {$R_c$};
\node at (2*\l, 0.8) {};
\node[draw, rounded corners, fit=(x.south) (x.west) (y.east) (y.north)] (C) {};
\node at (1*\l, 0.3) {$\textbf{k}$};
\end{tikzpicture}
\end{center}
  \end{minipage}
  \hfill
   \caption{Example of construction of the non-oriented graph modeling the random network of nanowires around four contacts and five wires. Each pair of nodes (separated by an edge associated to a contact resistance $R_c$) represents a contact, as the pair of nodes labeled $\mathbf{k}$. The edges not associated to a contact resistance correspond to linear resistance of nanowires, their weight depending on the distance between the contacts ($\rho$ is the wire resistance per unit length). The electric potential is thus well defined at each node.}
  \label{cas_contact_3_voisins_graphe_avec_Rc}
\end{figure}
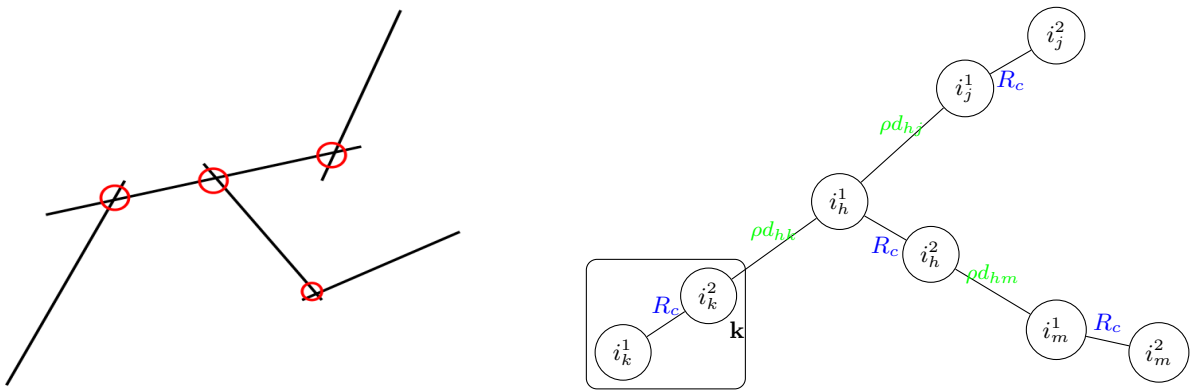

\begin{figure}[!h]
   \begin{minipage}[c]{0.40\linewidth}
   \centering
   \includegraphics[width=6cm,height=5cm]{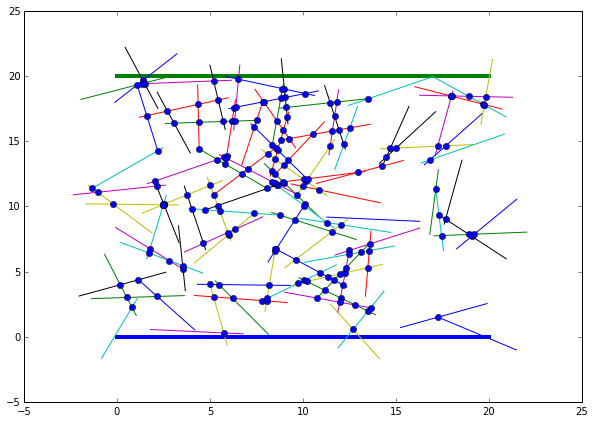}     
  \end{minipage}
  \hfill
\begin{minipage}[c]{0.40\linewidth}
   \centering
   \includegraphics[width=6cm,height=5cm]{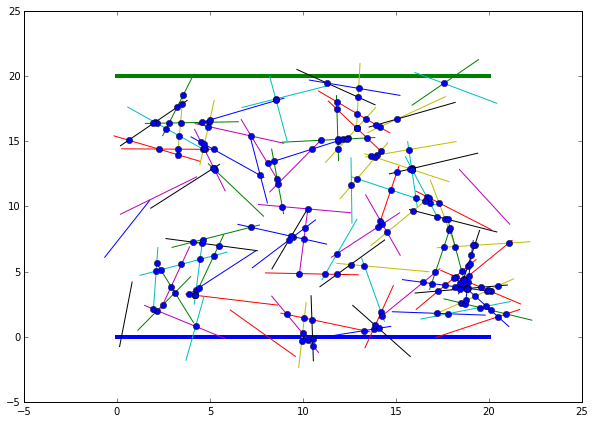}
  \end{minipage}
   \hfill
	\caption{Examples of randomly generated 2D percolating networks of carbon nanotubes (or any conducting nanowires), with parameters $N=100$, $l=L=20$, $l^{*}=5$. The electrodes are depicted by the blue and green bold horizontal lines. Percolation is reached in both cases. The calculated effective resistances are $\mathbf{R^{eq}_1=400 k\Omega}$ (with respective contributions of $230 k\Omega$ from all linear wire resistances, $164 k\Omega$ from all wire/wire contact resistances and $6 k\Omega$ from electrode/wire resistances) for the device on the left, and $\mathbf{R^{eq}_2=487 k\Omega}$ (with contributions of $232 k\Omega$ from all linear wire resistances, $249 k\Omega$ from all wire/wire contact resistances and $6 k\Omega$ from electrode/wire resistances) for the device on the right, taking a linear wire resistance per unit length $\rho =6 k\Omega / \mu m$ (from Ref. \cite{Hecht2006}), a nanowire/nanowire contact resistance $R_c=100 k\Omega$ (typical values for CNTs, see Ref. \cite{Nirmalraj2009}), and an electrode/nanowire contact resistance $R_{m,w}=10 k\Omega$.}
	\label{example_percolating_network_1}
\end{figure}

Two-dimensional random percolating networks of nanowires are generated by chosing randomly the positions of the centers of $N$ wires in the rectangle $[0,L] \times [0,l]$ representing the device surface ($L$ being the length of the channel and $l$ the width of the electrodes) and their angles with respect to the direction of the two electrodes (represented by the segment lines $[0,L] \times \left\lbrace 0 \right\rbrace $ and $[0,L] \times \left\lbrace l \right\rbrace $), following a uniform law in both cases. All the wires are assumed to have the same length $l^*$. All the positions of the contacts between two wires are found analytically (as the wires are assumed widthless) and associated to "internal" nodes of an underlying graph. The contacts which are on top of one of the electrodes are discarded as the current will flow preferentially from the metallic electrode to the nearest contact inside the device. Two examples of generated networks are depicted on Figure \ref{example_percolating_network_1}, with the detailed contributions of each type of resistance to the effective resistance (see section \ref{section_Results}). An example of construction of the underlying graph around 4 contacts is illustrated on Figure \ref{cas_contact_3_voisins_graphe_avec_Rc}. The "internal" nodes are first linked by edges of distance dependent resistance ($\rho d_{ij}$ where $\rho$ is the wire resistance per unit length and $d_{ij}$ the distance between two contacts indexed $i$ and $j$). The transport in individual wires is thus considered diffusive. Each of these nodes is then duplicated, and edges associated to contact resistances $R_c$ are added between each new pair of duplicated nodes. Finally, two other "external" nodes are added to account for the two electrodes, and linked to nearest contacts along the wires contacting them (with an edge associated to a resistance $R_{m,w}+\rho d$, where $R_{m,w}$ is the contact resistance between the metallic electrode and the wire, $d$ being the shortest distance between the electrode and the first contact along the wire). From the construction of the graph detailed above, $P$, the total number of nodes, is of order twice the number of contacts between nanowires.\\

The conductance matrix associated to this underlying graph is built accordingly, and allows to express Kirchhoff's laws, written at all the nodes, in a matrix form. We define the vector $\vec I$ of all the (algebraic) currents leaving the $P$ nodes :

\begin{equation}
\vec I =\begin{pmatrix}
I_1 \\
I_2 \\
... \\
I_{P-1} \\
I_{P}
\end{pmatrix} =\begin{pmatrix}
I \\
0\\
... \\
0 \\
-I
\end{pmatrix}
\end{equation}
which accounts for the injected current at the lower electrode (labeled by index $1$), flowing out at the higher electrode (labeled by index $P$). At all the other "internal" nodes, the current entering the node is equal to the current leaving the node. The entries of the vector $\vec V$ are the potentials at all the nodes of the graph (including the two nodes corresponding to the two electrodes) :
\begin{equation}
\vec V =\begin{pmatrix}
V_1 \\
V_2 \\
... \\
V_{P}
\end{pmatrix}
\end{equation}
$\vec I$ and $\vec V$ are related one to another by the conductance matrix $\mathcal{L}$ :
\begin{equation}
\fbox{$
\vec I = \mathcal{L} \vec V
$}
\label{I_Laplacian_V}
\end{equation}
Equation $\ref{I_Laplacian_V}$ simply corresponds to the Kirchhoff laws written at each node. The (symmetric positive) matrix $\mathcal{L}$ in equation \ref{I_Laplacian_V} is given by :
\begin{equation}
\mathcal{L}=\begin{pmatrix}
\sum_{j \neq 1} c_{1,j} & -c_{1,2} & ... & ... & ... & ... & ... & -c_{1,P} \\
-c_{1,2} & \sum_{j \neq 2} c_{2,j} & ... & ... & ... & ... & ... & -c_{2,P} \\
... & ...& ... & ... & ... &... & ... & ... \\
-c_{1,i} & ... & ... & \sum_{j \neq i} c_{i,j}  & ... & -c_{i,j} & ... & -c_{i,P}\\
... & ...& ... & ... &... & ... & ... & ... \\
... & ...& ... & ... &... & ... & ... & ... \\
-c_{1,P} & -c_{2,P} & ... & ... & ... & ... & ... & \sum_{j \neq P} c_{P,j} \\
\end{pmatrix}
\end{equation}
where $c_{i,j}=c_{j,i}$ is equal either to $\frac{1}{\rho d_{ij}}$ (edge of linear resistance), $\frac{1}{R_c}$ (edge of contact resistance), or $\frac{1}{\rho d_{1,i}+R_{m,w}}$ (edge between the metallic electrode and an internal node). We choose the same contact resistance to model all the junctions on the network (all the edges associated to contact resistances have thus the same conductance $\frac{1}{R_c}$), while the edges of nanowire linear pathways have a resistance proportional to the distance (contrary to the model of Ref. \cite{Kim2018}), \textit{i.e.} the transport in nanowires is assumed to be diffusive. It would be also possible to handle probabilistic distributions of contact and linear resistances.\\

In our model, each node is connected to a contact resistance edge, and one or two wire segments (one only in the case of a dead-end attached wire segment). Dead-end wire segments are not represented in the graph as they do not participate to the conduction. The degree of the nodes of the graph is thus either two or three for all inner nodes. The two nodes representing the two electrodes have a much higher connectivity, corresponding to the number of pathway departures or endings.\\

\paragraph*{Non-dimensionalization of the problem :\\}

We assume that the device is large enough so that the effective resistance has ohmic behavior, \textit{i.e.} is inversely proportional to the device width. Thus, we can choose a square device area ($L=l$) without loss of generality. The effective resistance $R^{eq}$ depends thus \textit{a priori} on six independent physical variables $\rho$, $R_c$, $R_{m,w}$, $L$, $l^*$ and $N$, so that there is a function $f$ satisfying $f(R^{eq},\rho,R_c,R_{m,w},L,l^*,N)=0$. Performing a dimensional analysis and applying Buckingham $\pi$ theorem leads to five independent non-dimensional variables only, which can be for instance taken as $\frac{R^{eq}}{R_c},\frac{\rho l^*}{R_c},\frac{R_{m,w}}{R_c},\frac{L}{l^*},N$ -- see Appendix $B$ for details. We can thus write the complete dependence of the non-dimensional effective resistance $\hat{R}^{eq}$ on both physical non-dimensional parameters $x=\frac{\rho l^*}{R_c}$ and $y=\frac{R_{m,w}}{R_c}$ and geometrical parameters as :
\begin{equation}
    \hat{R}^{eq}(x,y,N,\frac{L}{l^*})
\end{equation}
In order to obtain explicitly this form, we rewrite a non-dimensionalized version of equation \ref{I_Laplacian_V} :
\begin{equation}
\frac{\vec I}{I_0}=R_0 \mathcal{L} \frac{\vec{V}}{V_0} \Longleftrightarrow \vec{\hat{I}}=R_0 \mathcal{L} \vec{\hat{V}}
\end{equation}
where $I_0$ is a reference current, $V_0$ a reference potential, and $R_0=\frac{V_0}{I_0}$ a reference resistance. $\vec{\hat{V}}$ and $\vec{\hat{I}}$ are the non-dimensionnalized vectors of all node potentials and of all current leaving the nodes. The conductance matrix has coefficients of the following form, only :
\begin{equation}
\mathcal{L} \left(\left\lbrace\frac{1}{\rho d_{i,j}} \right\rbrace_{(i,j)}, \frac{1}{R_c}, \left\lbrace\frac{1}{\rho d_{1,i}+R_{m,w}} \right\rbrace_{i \in \mathcal{J}_1},\left\lbrace\frac{1}{\rho d_{P,k}+R_{m,w}} \right\rbrace_{k \in \mathcal{J}_2} \right)
\label{M*_dependances_bis}
\end{equation}
where $\mathcal{J}_1$ and $\mathcal{J}_2$ are the set of indexes of nodes linked to the lower and outer electrode respectively. We choose $l^{*}$, the length of a wire, as a reference distance and $R_0=R_c$, the contact resistance, as normalizing resistance. $R_c \mathcal{L}$ can be rewritten as a matrix $\hat{\mathcal{L}}$ with non-dimensional entries :
\begin{equation}
R_c \mathcal{L}  = \hat{\mathcal{L}} \left(\left\lbrace\frac{1}{\frac{\rho l^*}{R_c} \frac{d_{i,j}}{l^*}} \right\rbrace_{(i,j)}, 1, \left\lbrace\frac{1}{\frac{\rho l^*}{R_c} \frac{d_{1,i}}{l^*}+\frac{R_{m,w}}{R_c}} \right\rbrace_{i \in \mathcal{J}_1},\left\lbrace \frac{1}{\frac{\rho l^*}{R_c} \frac{d_{P,k}}{l^*}+\frac{R_{m,w}}{R_c}} \right\rbrace_{k \in \mathcal{J}_2} \right)
\label{M*_dependances_bis1}
\end{equation}

Let us introduce the non-dimensional parameters $x=\frac{\rho l^*}{R_c}$ and $y=\frac{R_{m,w}}{R_c}$. We therefore have rewritten the conductance matrix as :
\begin{small}
\begin{equation}
\fbox{$
\mathcal{L} (\rho,R_c,R_{m,w})=\frac{1}{R_c} \hat{\mathcal{L} } \left(x=\frac{\rho l^*}{R_c},y =\frac{R_{m,w}}{R_c} \right) =  \frac{1}{R_c} \hat{\mathcal{L}} \left(\left\lbrace\frac{1}{x \frac{d_{i,j}}{l^*}} \right\rbrace_{(i,j)}, 1, \left\lbrace\frac{1}{x \frac{d_{1,i}}{l^*}+y} \right\rbrace_{i \in \mathcal{J}_1},\left\lbrace \frac{1}{x \frac{d_{P,k}}{l^*}+y} \right\rbrace_{k \in \mathcal{J}_2} \right)
$}
\label{M*_dependances_bis2}
\end{equation}
\end{small}
The ratios $\frac{d_{hk}}{l^*}$ are geometrical features, and depend on the morphology of the randomly generated network only -- they are thus constant for a given fixed network. The ratio $x=\frac{\rho l^*}{R_c}$ of wire linear resistance $\rho l^*$ to wire/wire contact resistance $R_c$, plays as a parameter in the matrix, and so does the ratio $y$ of the contact resistance between electrode and nanowire to the junction resistance. Their values control the conduction regime, for instance dominated by contact resistance if $x \ll 1$. 

Solving the resistor network problem formulated as $\vec{\hat{I}}=\hat{\mathcal{L}} \vec{\hat{V}}$ is elementary for networks of moderate size. Several methods are possible to solve Kirchhoff laws, \textit{e.g.} thanks to iterative solvers \cite{Zezelj2012,Simoneau2013,Simoneau2015}. The details of the calculation methods are reported in Appendix $A$. The true resistance, expressed as a function of the physical parameters, is therefore : 
\begin{equation}
\fbox{$
R^{eq}(\rho,R_c,R_{m,w})=R_c \hat{R}^{eq}\left(x=\frac{\rho l^*}{R_c},y =\frac{R_{m,w}}{R_c} \right) 
=R_c \hat{R}^{eq}\left(x,y \right) $}
\label{expression_R_eq_function_R_eq_adimensionnelle}
\end{equation}
where $\hat{R}^{eq}$ is the non-dimensional effective resistance, in units of $R_c$, computed from the non-dimensional conductance matrix $\hat{\mathcal{L}}$. Non-dimensionalizing the problem has enabled us to reduce a problem dependent on three physical variables ($\rho$, $R_c$ and $R_{m,w}$) to a problem function of two variables ($x$ and $y$) only. The dependence of the resistance on the random network generating process, on its structure and geometry (number of wires $N$ or normalized density $\frac{N}{\left( \frac{L}{l^*} \right)^2}$, dimensions $L \times l$ of the device area) is implicit through the geometrical (random) factors $\frac{d_{hk}}{l^*}$ factoring $x$ and $y$ in the matrix $\hat{\mathcal{L}}$ and via the size of the matrix $\mathcal{L}$. This matrix dimension is of order twice the number of contacts. The latter depends directly on the number of wires $N$ and on the aspect ratio $\frac{L}{l^*}$ (from Ref. \cite{Heitz2011}, it can be estimated as $\frac{0.2027 \pi}{2 \left( \frac{L}{l^*} \right)^2}N^2$).\\
In the following, we perform numerical simulations to estimate an analytical expression of $\hat{R}^{eq}(x,y)$, focusing on the functional dependence on physical parameters $x$ and $y$ -- the associated geometrical factors being the results of fitting and averages.

\section{Results}
\label{section_Results}

The functional dependence of the effective resistance on the physical parameters ($\rho$, $R_c$, $R_{m,w}$, \textit{i.e.} $x$ and $y$ in terms on non-dimensional parameters) has been first investigated for fixed percolating networks of $N$ nanowires, at a given aspect ratio $\frac{L}{l^*}$. Then, for the sake of generality and applicability of the model to real devices -- of random, unpredictable geometry --, this functional dependence on the resistance of the individual components of the network ($\rho$, $R_c$, $R_{m,w}$) was generalized for average networks by averaging, at a given aspect ratio $\frac{L}{l^*}$, over several realizations of random networks made up of $N$ nanowires.\\

\paragraph*{At fixed random network :\\}

Fixing a percolating network allows one to fix the geometry and to have the physical parameters (resistances) vary, through the variation of $x$ and $y$, without the variability due to the randomness of the network construction. For a given fixed network, the effective non-dimensional resistance $\hat{R}^{eq}(x,y)$ obtained by solving the linear system $\vec{\hat{I}}=\hat{\mathcal{L}}(x,y) \vec{\hat{V}}$ is in the most general case a rational fraction of $x$ and $y$ (see Appendix $C$ for the details of the calculations). 
Here, we show numerically that $\hat{R}^{eq}(x,y)$ can be reduced to a very good level of accuracy, for any fixed network of nanowires above percolation, to a simple sum a linear terms in $x$ and $y$, plus a constant : $Ax+Cy+B$ (over the physically relevant ranges of values of $x$ and $y$). This formula appears to be a very good approximation at any density above the percolation level.\\

First, a random network of nanowires is generated and its associated conductance matrix computed. The effective resistance is computed for values of $x$ such that the ratio $\frac{\rho \times (1 \mu m)}{R_c}= \frac{1 \mu m}{l^*}x$ varies over the range 0.01 to 1 (with steps of $0.01$) and such that $y=\frac{R_{m,w}}{R_c}$ varies over the range 0.1 to 1 (with steps of $0.1$), \textit{i.e.} for a total of 1000 different values of $(x,y)$. This enables to span all the conduction regimes -- from junction resistances dominated to linear resistance dominated, for instance. The function $(x,y) \mapsto \hat{R}^{eq}(x,y)$ is then fitted to the obtained data points by linear regression. The geometrical parameters $N,\frac{L}{l^*}$ are thus fixed. \\
This procedure can be repeated for different values of $N$ and $\frac{L}{l^*}$ -- fixing a given generated network, for each value of $\left(N,\frac{L}{l^*} \right)$. We obtain the following form, from least-squares fitting of $(x,y) \mapsto \hat{R}^{eq}(x,y)$ to the computed data points, at each fixed geometry  $\left(N,\frac{L}{l^*} \right)$ : 
\begin{equation}
\hat{R}^{eq}(x,y) \simeq A \left(N,\frac{L}{l^*} \right)x+C\left(N,\frac{L}{l^*} \right)y +B\left(N\frac{L}{l^*}\right) 
\label{R_eq_function_x_y_case}
\end{equation}
The result \ref{R_eq_function_x_y_case} is found to be valid for any network above the percolation threshold (with varying approximation errors on parameters $A\left(N,\frac{L}{l^*}\right)$, $B\left(N,\frac{L}{l^*}\right)$ and $C\left(N,\frac{L}{l^*}\right)$ depending on the geometry and the distance with respect to percolation). Examples of coefficients $A\left(N,\frac{L}{l^*}\right)$, $B\left(N,\frac{L}{l^*} \right)$ and $C\left(N,\frac{L}{l^*} \right)$, for fixed nanowire networks of different densities (for a same device aspect ratio $\frac{L}{l^*}=4$), are given table \ref{tableau_A_B_fonction_N_reseaux_fixes}, as well as fitting errors. It can be seen from the results displayed in table \ref{tableau_A_B_fonction_N_reseaux_fixes} that equation \ref{R_eq_function_x_y_case} holds to a very high degree of accuracy, given the very low standard deviations on the fitted geometrical parameters $A$, $B$ and $C$.\\

In physical units, from equation \ref{R_eq_function_x_y_case}, the true resistance of a fixed given network is :
\begin{equation}
\fbox{$
R^{eq}(\rho,R_c,R_{m,w})= R_c \hat{R}^{eq}(x,y) \simeq A\left(N,\frac{L}{l^*}\right) \rho l^*+ B\left(N,\frac{L}{l^*}\right) R_c + C\left(N,\frac{L}{l^*}\right) R_{m,w} 
$}
\label{R_eq_y=0}
\end{equation}
This decomposition of the effective resistance into its three different contributions (of different physical origin) is given for the random networks of Figure \ref{example_percolating_network_1}. The coefficients $A\left(N,\frac{L}{l^*}\right)$, $B\left(N,\frac{L}{l^*}\right)$ and $C\left(N,\frac{L}{l^*}\right)$ are purely geometrical factors. They could be interpreted as the ratio of the mean number of linear portions (or respectively of nanowire/nanowire junctions, or electrode/nanowire contacts) per "typical" percolation pathway, to the number of such typical pathways (in a picture of parallel identical pathways). A very simple interpretation can also be a model with three resistances in series, each one accounting for the contributions of all the resistances of the network of a certain kind (linear, wire/wire junction, or electrode/wire contact), rescaled by a geometrical factor dependent on the number of wires and aspect ratio only. \\

\begin{table}[!h]\centering
\begin{tabular}{|>{\centering}m{1.8cm}|>{\centering}m{2.3cm}|>{\centering}m{2cm}|>{\centering}m{1.5cm}|>{\centering}m{2.cm}|>{\centering}m{1.5cm}|>{\centering}m{2.cm}|>{\centering}m{1.5cm}|}
\hline
 Number of nanowires $N$ & Distance to percolation $\frac{N}{N_{cr}}$ & $A\left(N,\frac{L}{l^*}=4 \right)$ & $\epsilon_{A(N)}$ & $B\left(N,\frac{L}{l^*}=4 \right)$ & $\epsilon_{B(N)}$  &  $C\left(N,\frac{L}{l^*}=4 \right)$ & $\epsilon_{C(N)}$ \tabularnewline
 \hline
 150 & 1.66 & 0.818 & 2.00 10$^{-4}$ & 1.153 & 1.00 10$^{-3}$ & 0.195 & 1.20 10$^{-3}$  \tabularnewline
\hline
200 & 2.22 &  0.435 & 2.00 10$^{-4}$ & 0.482 & 7.00 10$^{-4}$ & 0.196 & 8.00 10$^{-4}$ \tabularnewline
\hline
250 & 2.77 & 0.267 & 1.00 10$^{-4}$ & 0.260 & 4.00 10$^{-4}$ &  0.118 & 5.00 10$^{-4}$ \tabularnewline
\hline
300 & 3.33 & 0.178 & 7.00 10$^{-5}$ & 0.145 & 3.00 10$^{-4}$ & 8.21 10$^{-2}$ & 4.00 10$^{-4}$ \tabularnewline
\hline
350 & 3.88 & 0.149 & 6.00 10$^{-5}$ & 0.101 & 3.00 10$^{-4}$  & 8.28 10$^{-2}$ & 3.00 10$^{-4}$ \tabularnewline
\hline
400 & 4.43 & 0.126 & 5.00 10$^{-5}$  & 8.27 10$^{-2}$ & 2.00 10$^{-4}$ & 5.73 10$^{-2}$ & 3.00 10$^{-4}$  \tabularnewline
\hline
450 &  4.99 & 0.116 &  5.00 10$^{-5}$ & 6.81 10$^{-2}$ &  2.00 10$^{-4}$  & 7.19 10$^{-2}$ & 2.00 10$^{-4}$ \tabularnewline
\hline
\end{tabular}
\caption{Non-dimensional coefficients $A\left(N,\frac{L}{l^*}=4 \right)$, $B\left(N,\frac{L}{l^*}=4 \right)$ and $C\left(N,\frac{L}{l^*}=4 \right)$ as a function of $N$, for a fixed single realization of the network for each $N$, and a geometry $l=L=20$, $l^{*}=5$ (aspect ratio $\frac{L}{l^{*}}=4$), plotted Figure \ref{evolution_A_B_function_N_aspect_ratio_4_log_log} (red points). The critical number of wires to reach percolation $N_{cr} \approx 5.64 \left( \frac{L}{l^{*}} \right)^2$ in two-dimensional stick networks is defined such that the probability of percolation is exactly 0.5 at $N=N_{cr}$ \cite{Li2009a}. The values of $A$, $B$ and $C$ are derived from least-squares fitting of the simulated function $(x,y) \mapsto \hat{R}^{eq}(x,y)$ to a functional form $(x,y) \mapsto Ax+B+Cy$. The chosen grid of points is $\left\lbrace (x_i,y_j) \right\rbrace_{i,j}$ where $x_i =\frac{l^*}{1 \mu m} (1+i)0.01$, $y_j=(1+j)0.1$ and $(i,j) \in [\![0,99 ]\!] \times [\![0,9 ]\!]$. $\epsilon_{A(N)}$ and $\epsilon_{B(N)}$ and $\epsilon_{C(N)}$ are the approximation errors on these three coefficients from the linear regression (square root of the diagonal terms of the estimated parameters covariance matrix).}
\label{tableau_A_B_fonction_N_reseaux_fixes}
\end{table}

\paragraph*{Averaging over random networks :\\}

Real random percolating networks of nanowires do not have a predictable geometry. Rather, they can be viewed as representatives of the class of all possible random percolating networks of nanowires at a given density (which is known or at least estimated experimentally). The detailed structure (\textit{i.e.} the positions of all the wires, angles, and positions of the contacts) of a real random network of nanowires -- of given density -- is difficult to characterize from imaging techniques, specially at very high density, well above the percolation threshold. Capturing the actual precise distribution of wires from an SEM analysis of a real network was nonetheless done in Ref. \cite{Rocha2015}, for very low density networks of silver nanowires. As the reproducibility of the resistance of a given network increases with its density, for high density networks, a good estimation of the geometrical factors $A$, $B$ and $C$ governing the effective resistance would be given by the coefficients averages over some realizations of random networks of same density.\\

For this purpose, we averaged, for each value of the physical variables $(x,y)$, the computed values of resistances for several random networks made up of $N$ nanowires (with $\frac{L}{l^*}$ fixed). We obtained, after fitting, the same functional dependence (on physical parameters) for the average effective resistance of random percolating networks made up of $N$ nanowires (at a fixed geometry $\frac{L}{l^*}$), this time with average coefficients :
\begin{equation}
    \overline{\hat{R}^{eq}}(x,y)=\overline{A\left(N,\frac{L}{l^*}\right)} x +\overline{C\left(N,\frac{L}{l^*}\right)}y + \overline{B\left(N,\frac{L}{l^*}\right)}
    \label{R_eq_x_y_averaged}
\end{equation}
in terms of non-dimensional effective resistance and :
\begin{equation}
\overline{R^{eq}}(\rho,R_c,R_{m,w})= R_c \overline{\hat{R}^{eq}}(x,y) =\overline{A\left(N,\frac{L}{l^*}\right)} \rho l^* +\overline{B\left(N,\frac{L}{l^*}\right)} R_c + \overline{C\left(N,\frac{L}{l^*}\right)} R_{m,w}
\label{R_eq_physical_parameters_averaged}
\end{equation}
in physical units. \\

\begin{figure}[!h]\centering
   \includegraphics[width=9cm,height=8cm]{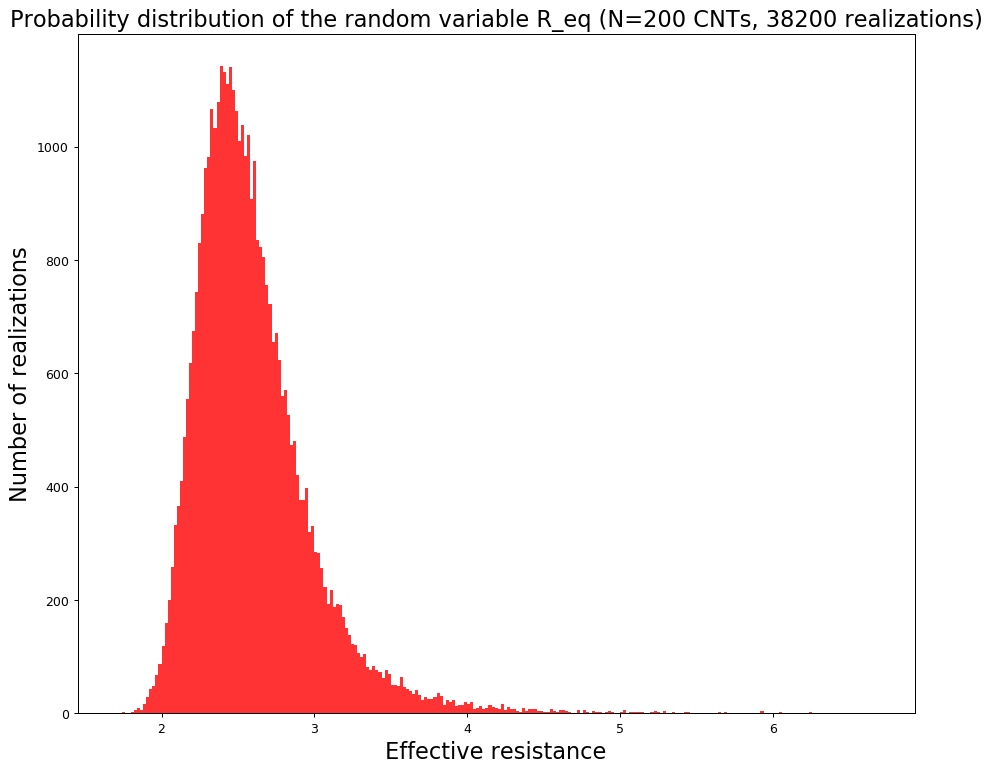}
	\caption{Effective resistance (in units of the contact resistance $R_c$) probability distribution (histogram constructed from 38200 realizations) of the random variable $\omega \in \Omega_{N=200,\frac{L}{l^*}=4} \longmapsto R^{eq}\left( \omega \right)$ \textit{i.e.} the effective resistance of random networks made up of 200 nanowires, with a device aspect ratio $\frac{L}{l^*}=4$, all the parameters being fixed ($x=1$, $y=1$ \textit{i.e.} $\rho l^{*}=R_{m,w}=R_c$).}
	\label{histogram_R_eq_variable_aleatoire_N_200CNTs}
\end{figure}

Let $\Omega_{N,\frac{L}{l^*}}$ be the set of possible realizations of random networks of $N$ nanowires (with a geometry characterized by the aspect ratio $\frac{L}{l^*}$) and $\omega \in \Omega_{N,\frac{L}{l^*}}$ denote a particular realization. Expression \ref{R_eq_x_y_averaged} amounts to approximate, at each fixed value of $(x,y)$ and fixed values of $N$ and $\frac{L}{l^*}$, the effective resistance $\omega \longmapsto R^{eq}(\omega)$ (seen as a random variable from $\Omega_{N,\frac{L}{l^*}}$ to $\mathbb{R}$) by its average $\overline{R^{eq}}$. The effective resistance of random networks of nanowires, all parameters being fixed, is indeed a well-behaved unimodal random variable, which can be fairly well approximated by its average value and standard deviation. We justified this assumption by calculating probability distributions of the network effective resistance, all parameters (geometrical and physical) being fixed. It yielded a trend depicted on Figure \ref{histogram_R_eq_variable_aleatoire_N_200CNTs} -- which had already been reported in the context of small-world resistor networks \cite{Korniss2006}. 
Varying the number of nanowires $N$ and the aspect ratio $\frac{L}{l^*}$ only changes the mean resistance and the standard deviation but does not change the shape of this law. Incidentally, this law gives an indication of the experimental non-reproducibility of the resistance that has to be expected when fabricating random nanowire networks.\\

The expression :
\begin{equation}
\overline{R^{eq}}(\rho,R_c,R_{m,w}=0)= R_c \overline{\hat{R}^{eq}}(x,y=0) =\overline{A\left(N,\frac{L}{l^*}\right)} \rho l^* +\overline{B\left(N,\frac{L}{l^*}\right)} R_c
\label{R_eq_x_y=0_averaged}
\end{equation}
at $R_{m,w}=0$ is reminiscent of the empirical expression of Zezelj \textit{and al.} \cite{Zezelj2012} successfully introduced to explain  the high-density behavior of the local conductivity exponent depending on the junction resistance to wire resistance ratio (no electrode/wire contact resistance was included). The geometrical factors proposed in Ref. \cite{Zezelj2012} are $\overline{A(N)} \varpropto \frac{1}{N}$ and $\overline{B(N)} \varpropto \frac{1}{N^2}$, for dense networks, well above percolation, and at fixed aspect ratio $\frac{L}{l^*}=20$. \\

\begin{figure}[!h]
   \begin{minipage}{0.45\linewidth}
   \centering
   \includegraphics[width=6cm,height=5cm]{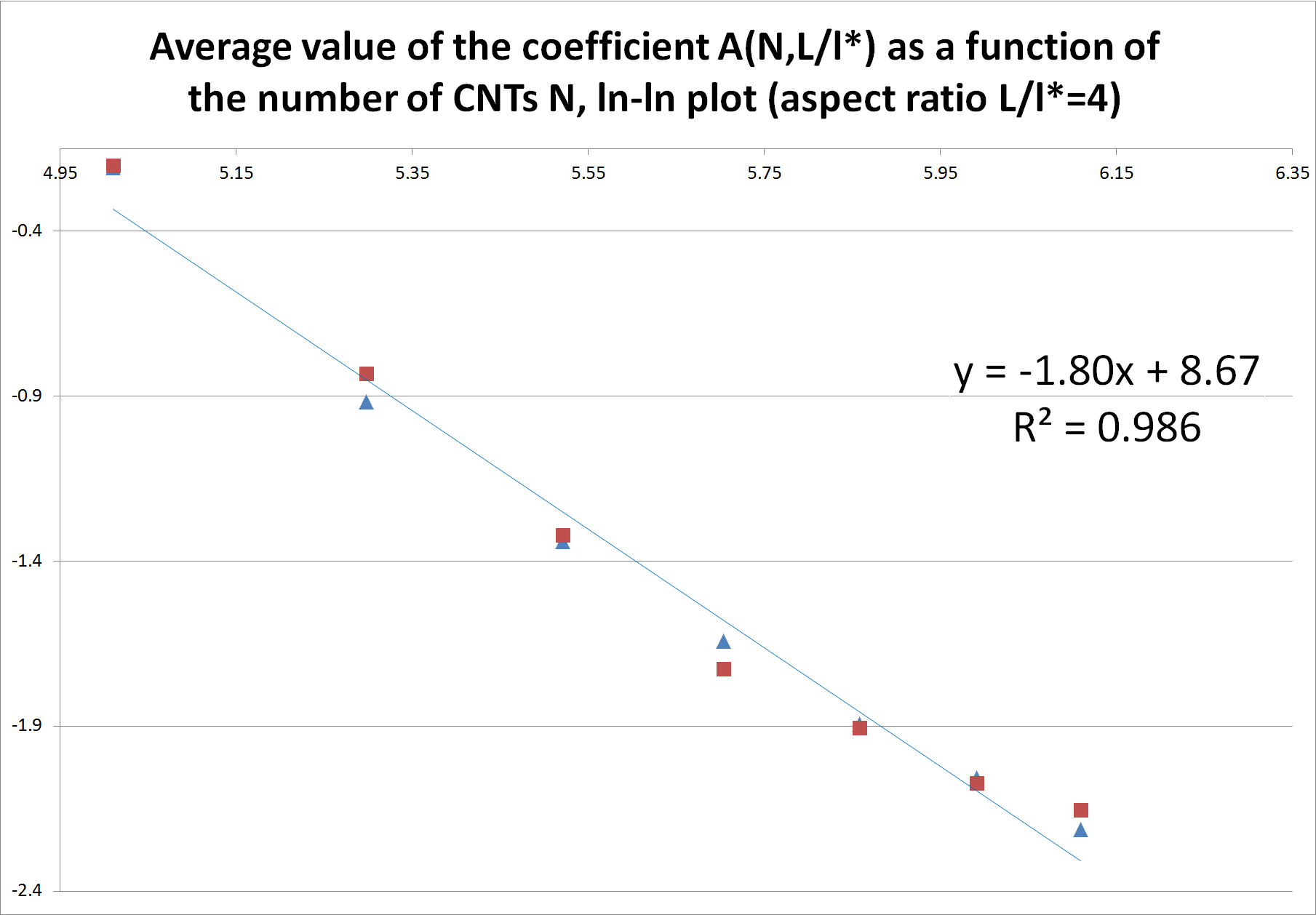}
  \end{minipage}
  \hfill
 \begin{minipage}{0.45\linewidth}
  \centering
   \includegraphics[width=6cm,height=5cm]{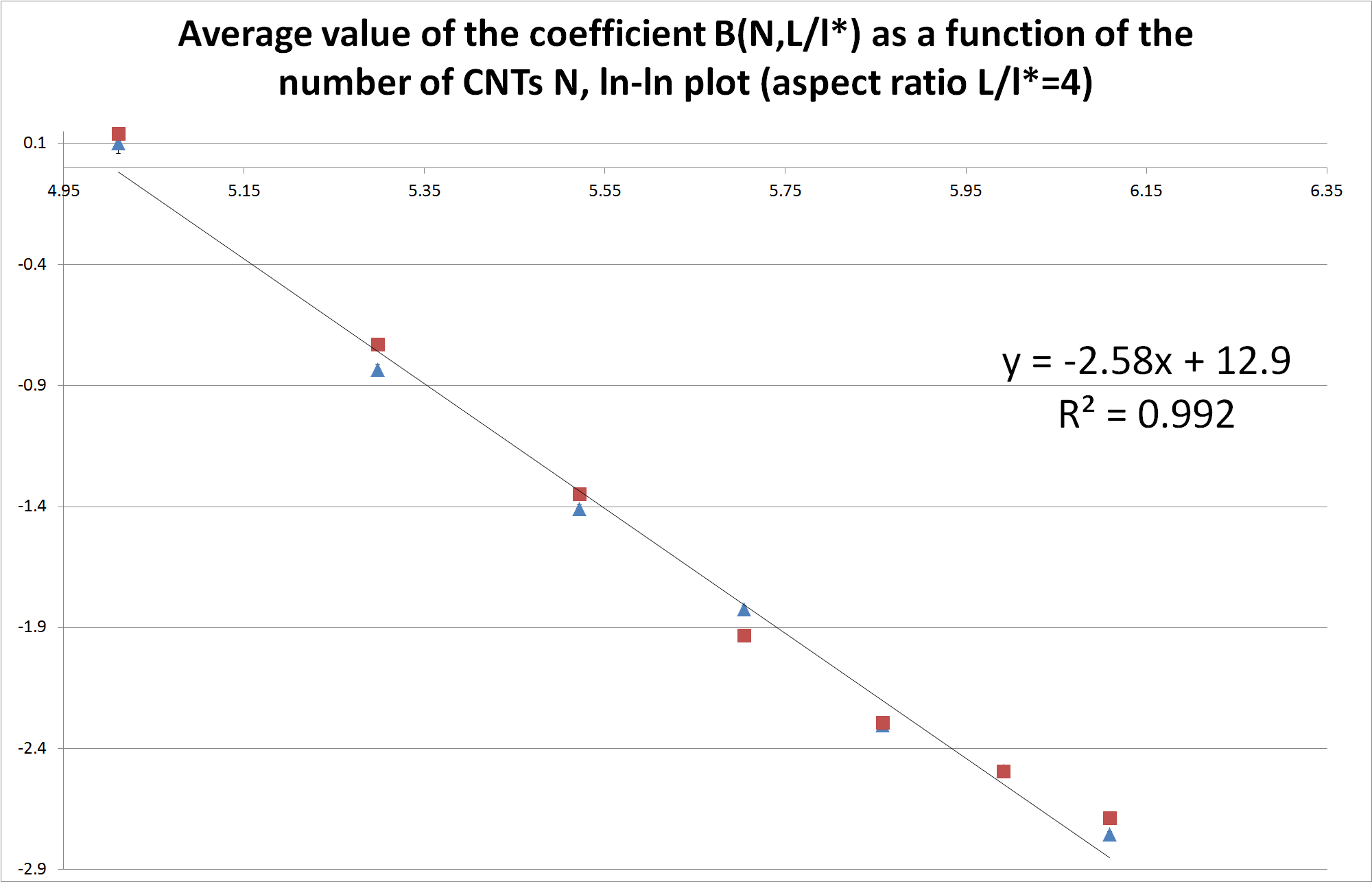}
 \end{minipage}
 \begin{minipage}{0.45\linewidth}
  \centering
   \includegraphics[width=6cm,height=5cm]{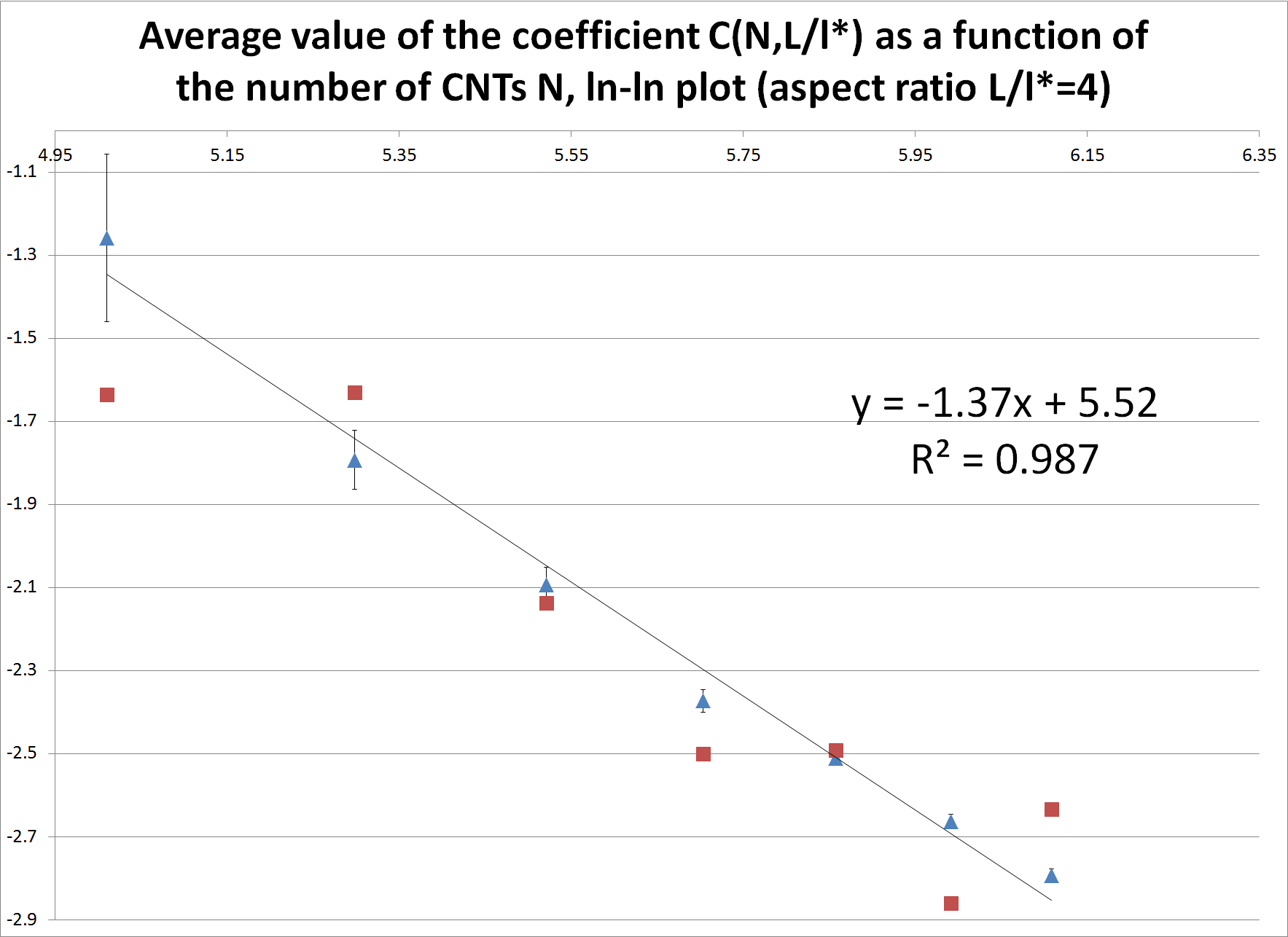}
 \end{minipage}
 \caption{Evolution of the average coefficients $\overline{A\left(N,\frac{L}{l^*}=4 \right)}$, $\overline{B\left(N,\frac{L}{l^*}=4 \right)}$ and $\overline{C\left(N,\frac{L}{l^*}=4 \right)}$ of Table \ref{evolution_A_B_C_function_N_aspect_ratio_4} (blue points) and of the coefficients $A\left(N,\frac{L}{l^*}=4 \right)$, $B\left(N,\frac{L}{l^*}=4 \right)$ and $C\left(N,\frac{L}{l^*}=4 \right)$ of the single, fixed networks of Table \ref{tableau_A_B_fonction_N_reseaux_fixes} (red points), as a function of $N$ ($L=l=20$, $l^{*}=5$, aspect ratio $\frac{L}{l^{*}}=4$), log-log plot (with standard deviations). The linear least-squares fit is done relatively to the average coefficients (blue points).}
 \label{evolution_A_B_function_N_aspect_ratio_4_log_log}
 \end{figure}

The values found for the average coefficients $\overline{A\left(N,\frac{L}{l^*} \right)}$, $\overline{B\left(N,\frac{L}{l^*} \right)}$ and $\overline{C\left(N,\frac{L}{l^*} \right)}$ for an aspect ratio $\frac{L}{l^*}=4$ are reported table \ref{evolution_A_B_C_function_N_aspect_ratio_4} as a function of the number of wires $N$, and plotted on Figure \ref{evolution_A_B_function_N_aspect_ratio_4_log_log}. These three geometrical coefficients decrease with increasing number of wires following a power-law dependence, typical of percolation : $\overline{A\left(N,\frac{L}{l^*}=4 \right)} \varpropto \frac{1}{N^{\gamma}}$, $\overline{B\left(N,\frac{L}{l^*}=4 \right)} \varpropto \frac{1}{N^{\beta}}$ and $\overline{C\left(N,\frac{L}{l^*}=4 \right)} \varpropto \frac{1}{N^{\delta}}$ with critical exponents $\gamma \approx 1.8 $, $\beta \approx 2.6$ and $\delta \approx 1.4$. We note that these exponents are different from those of Ref. \cite{Zezelj2012}. \\

\begin{table}[!h]\centering
\begin{tabular}{|>{\centering}m{1.8cm}|>{\centering}m{2.3cm}|>{\centering}m{2cm}|>{\centering}m{1.5cm}|>{\centering}m{2.cm}|>{\centering}m{1.5cm}|>{\centering}m{2.cm}|>{\centering}m{1.5cm}|}
\hline
 Number of nanowires (N) & Distance to percolation $\frac{N}{N_{cr}}$ & $\overline{A\left(N,\frac{L}{l^*}=4 \right)}$ & $\sigma_{\overline{A(N)}}$ & $\overline{B\left(N,\frac{L}{l^*}=4 \right)}$ & $\sigma_{\overline{B(N)}}$ & $\overline{C\left(N,\frac{L}{l^*}=4 \right)}$ & $\sigma_{\overline{C(N)}}$  \tabularnewline
 \hline
 150 & 1.66 & 0.812 & 1.14 10$^{-2}$ & 1.11 & 4.58 10$^{-2}$ & 0.284 & 5.74 10$^{-2}$  \tabularnewline
\hline
200 & 2.22 &  0.400 & 2.30 10$^{-3}$ & 0.435 & 9.40 10$^{-3}$ & 0.167 & 1.18 10$^{-2}$  \tabularnewline
\hline
250 & 2.77 & 0.262 & 1.00 10$^{-3}$ & 0.245 & 4.10 10$^{-3}$ & 0.123 & 5.10 10$^{-3}$ \tabularnewline
\hline
300 & 3.33 & 0.193 & 5.00 10$^{-4}$ & 0.162 & 2.10 10$^{-3}$ & 9.32 10$^{-2}$ & 2.60 10$^{-3}$ \tabularnewline
\hline
350 & 3.88 & 0.151 & 3.00 10$^{-4}$ & 0.100 & 1.10 10$^{-3}$ & 8.13 10$^{-2}$ &  1.40 10$^{-3}$ \tabularnewline
\hline
400 & 4.43 & 0.128 & 2.00 10$^{-4}$ & 8.27 10$^{-2}$ & 9.00 10$^{-4}$ &  6.98 10$^{-2}$ &  1.10 10$^{-3}$ \tabularnewline
\hline
450 &  4.99 & 0.109 & 2.00 10$^{-4}$ & 6.36 10$^{-2}$ & 7.00 10$^{-4}$ & 6.13 10$^{-2}$ & 9.00 10$^{-4}$ \tabularnewline
\hline
\end{tabular}
\caption{Average coefficients $\overline{A\left(N,\frac{L}{l^*}=4 \right)}$, $\overline{B\left(N,\frac{L}{l^*}=4 \right)}$ and $\overline{C\left(N,\frac{L}{l^*}=4 \right)}$ of the law $(x,y) \mapsto \hat{R^{eq}}(x,y)$ as a function of $N$, for a geometry $l=L=20$, $l^{*}=5$ (aspect ratio $\frac{L}{l^{*}}=4$). The values of $\overline{A}$, $\overline{B}$ and $\overline{C}$ are derived from least-squares fitting of the simulated function $(x,y) \mapsto \overline{\hat{R}^{eq}}(x,y)$ (average at each value of $(x,y)$ of the random realizations) to a functional form $(x,y) \mapsto Ax+B+Cy$. The grid chosen $\left\lbrace (x_i,y_j) \right\rbrace_{i,j}$ is the same as for fixed networks (see Table \ref{tableau_A_B_fonction_N_reseaux_fixes}). At each value of $(x,y)$, the effective resistance is averaged over at least 5 (and at most 15) random realizations of percolating networks of $N$ tubes (with a target standard deviation of 10 \% on the value of the computed resistance, not always reached at small densities). Standard deviations $\sigma_{\overline{A(N)}}$, $\sigma_{\overline{B(N)}}$ and $\sigma_{\overline{C(N)}}$ are given by the square root of the diagonal terms of the covariance matrix on the parameters estimated by linear regression. }
\label{evolution_A_B_C_function_N_aspect_ratio_4}
\end{table}

\newpage
As can be observed comparing Tables \ref{tableau_A_B_fonction_N_reseaux_fixes} and \ref{evolution_A_B_C_function_N_aspect_ratio_4}, those average coefficients $\overline{A\left(N,\frac{L}{l^*}\right)}$, $\overline{B\left(N,\frac{L}{l^*}\right)}$ and $\overline{C\left(N,\frac{L}{l^*}\right)}$ are at sufficiently high density ($\approx 3$ times the percolation threshold) very close to the coefficients $A\left(N,\frac{L}{l^*}\right)$, $B\left(N,\frac{L}{l^*}\right)$ and $C\left(N,\frac{L}{l^*}\right)$ obtained for particular realizations of random percolating networks (Table \ref{tableau_A_B_fonction_N_reseaux_fixes}). In other words, a random network of $N$ nanowires is increasingly representative, when $N$ increases, of the whole class of random networks made up of $N$ nanowires (at given fixed aspect ratio $\frac{L}{l^*}$), and fewer random realizations are needed to estimate the average coefficients $\overline{A\left(N,\frac{L}{l^*}\right)}$, $\overline{B\left(N,\frac{L}{l^*}\right)}$ and $\overline{C\left(N,\frac{L}{l^*}\right)}$ than at low densities (in other words, the width of the distribution of the random variable depicted Figure \ref{histogram_R_eq_variable_aleatoire_N_200CNTs} decreases with increasing density).\\

\section{Applications to the understanding and design optimization of nanowire network thin films and sensors}

The very simple form (\ref{R_eq_physical_parameters_averaged}) found for the effective resistance allows to separate clearly the dependence on physical and geometrical parameters. This feature can be of upmost importance for the design or understanding of thin films or sensors based on random percolating networks of nanowires. The origin of the measured resistance, and sensitivity, can be decomposed into three different types of resistances (of different physical origin) and controlled thanks to geometrical and structural features (density of wires and aspect ratio).\\

For instance, from the results obtained above for the average geometrical coefficients, at aspect ratio $\frac{L}{l^*}=4$ (Table \ref{evolution_A_B_C_function_N_aspect_ratio_4}), the ratio $\frac{\overline{A\left(N,\frac{L}{l^*}\right)}}{\overline{B\left(N,\frac{L}{l^*}\right)}}$ decreases following a power law as a function of the density, given the different critical exponents $\gamma$ and $\beta$ for $\overline{A\left(N,\frac{L}{l^*}\right)}$ and $\overline{B\left(N,\frac{L}{l^*}\right)}$ respectively :
\begin{equation}
    \frac{\overline{B\left(N,\frac{L}{l^*}=4\right) }R_c}{\overline{A\left(N,\frac{L}{l^*}=4\right)} \left( \rho l^* \right)} \varpropto \frac{R_c}{\rho l^*} \frac{1}{N^{0.8}}  \mathop{\longrightarrow}_{N \to +\infty } 0
    \label{relative_contributions_Rc_R_linear}
\end{equation}
Thus, the contribution to the whole effective resistance of the contact resistances (numerator in equation \ref{relative_contributions_Rc_R_linear}) decreases in proportion, relatively to the contribution of nanowire linear resistances (denominator in equation \ref{relative_contributions_Rc_R_linear}), when the density increases, at this fixed aspect ratio. Still, the prefactor $\frac{R_c}{\rho l^*}$ is controlled by the ratio of single contact to single wire resistance, and can take quite high values. From a sensor perspective, this makes a sensor based on 2D random percolating of nanowires increasingly sensitive to variations of linear resistances, compared to variations of contact resistances, when the density increases, at fixed aspect ratio. \\
A similar analysis gives that the contribution to the whole effective resistance of the electrode/wire contact resistances (numerator in equation \ref{relative_contributions_R_electrode_R_linear}) increases in proportion, relatively to the contribution of nanowire linear resistances (denominator in equation \ref{relative_contributions_R_electrode_R_linear}), when the density increases :
\begin{equation}
    \frac{\overline{C\left(N,\frac{L}{l^*}=4\right) }R_{m,w}}{\overline{A\left(N,\frac{L}{l^*}=4\right)} \left( \rho l^* \right)} \varpropto \frac{R_{m,w}}{\rho l^*} N^{0.4} \mathop{\longrightarrow}_{N \to +\infty } +\infty
    \label{relative_contributions_R_electrode_R_linear}
\end{equation}
From a sensor perspective \cite{Michelis2015}, optimization of device sensitivities to particular types of resistances (\textit{e.g.} of linear vs. contact origin) can be addressed, tuning either the density of nanowires in the networks, or the size of the device (aspect ratio). Indeed, the geometrical coefficients $\overline{A\left(N,\frac{L}{l^*}\right)}$, $\overline{B\left(N,\frac{L}{l^*}\right)}$ and $\overline{C\left(N,\frac{L}{l^*}\right)}$ govern the relative contributions of linear resistance, wire/wire contact resistance, and electrode/wire contact resistance, respectively, to the total effective resistance.\\

Given the unveiled additive nature for the effective resistance of random percolating networks of nanowires, another possible application is to estimate the values of unit base (single nanowire, single contact) resistances $\rho$, $R_c$ and $R_{m,w}$ (as done recently by Forr\'o \textit{and al.} \cite{Forro2018}). Let us indeed consider at least three nanowire networks devices, made up of the same nanowires (whose local intrinsic properties, as linear or contact resistances, are thus the same) but of different densities and (or) aspect ratios. Let us assume that the effective resistances of these devices are simultaneously measured. Geometrical coefficients $A\left(N_i,\frac{L_i}{l^*}\right)$, $B\left(N_i,\frac{L_i}{l^*}\right)$, $C\left(N_i,\frac{L_i}{l^*}\right)$ are estimated numerically either for the virtual 'twin' networks of the real networks -- in the case that the actual precise distribution of wires in the real network can be captured from an SEM analysis, as in Ref. \cite{Rocha2015} --, or from the average coefficients at the corresponding densities and aspect ratios. From these experimentally measured resistances, and numerically estimated geometrical coefficients, it is possible to recover the values of the elementary contact resistance between two wires, the wire resistance per unit length and the electrode/nanowire contact resistance (three unknowns, and at least three independent equations). The same idea would apply in a sensor perspective when resistance variations $\left( \delta R^{eq} \right)_1$, $\left( \delta R^{eq} \right)_2$ and $\left( \delta R^{eq} \right)_3$ are measured simultaneously for the same sensitive element deposited between pairs of electrodes, but for (at least) three channel of varying lengths (or varying densities). In that case the elementary variations of the physical parameters, $\delta \rho$, $\delta R_c$ and $\delta R_{m,w}$, that gave rise to the measured macroscopic resistance variation, could be recovered. The elementary resistances, or resistance variations, are derived by inverting a linear system of equations, involving the numerically estimated geometrical coefficients. As the inversion amplifies uncertainties, for the uncertainty on the estimated resistances to be low enough, the geometrical parameters have to be estimated with very high precision -- and also the density and mean length of the wires be known with high precision. 

\section{Conclusion}

Thanks to Monte-Carlo numerical simulations, we have derived an accurate closed-form approximation of the functional dependence of the effective resistance of two-dimensional random percolating networks of widthless, stick nanowires, on physical parameters (wire resistance, wire/wire contact resistance and metallic electrode/wire contact resistance). The influence of geometrical parameters (number of wires, aspect ratio) has been identified as modulating the contributions of each type of resistance to the whole, effective resistance. The expression found for the effective resistance, much simpler than the expressions previously reported, is found to be applicable on the whole range of densities, in the relevant ranges for physical parameters. The simple additive nature of the effective resistance, with respect to the three main types of resistances (of different origins), opens the way to numerous applications for nanowire networks understanding and device optimization, in particular in the field of sensors.\\

The influence of other improved (\textit{i.e.} closer to real networks) modelling choices, such as the dispersion in the length distribution of the generated wires -- or also wire waviness, wire width -- on the results found here for the effective resistance, have not been discussed. Arguably, only the geometrical coefficients $A(N,\frac{L}{l^*})$, $B(N,\frac{L}{l^*})$ and $C(N,\frac{L}{l^*})$ would be changed (possibly leading to different critical exponents) by this refinements, but not the functional form with respect to the physical parameters (resistances) of equation \ref{R_eq_physical_parameters_averaged}. This robustness on the random generating procedure remains to be checked in future works.\\ 
The generalization of this framework to the modelling of 3D networks would be very interesting to get closer to real carbon nanotubes percolating networks at moderate to high densities. It would be also captured by a graph with nodes of well-defined potentials, with a possibly increased connectivity of the nodes for cases of two superposed wire/wire contacts in adjacent layers. \\
Finally, studying the precise current distribution in the network, and \textit{e.g.} the current carrying fraction of the wires, as done for instance in Ref. \cite{Zezelj2012,Kumar2017}, as a function of varying physical parameters (\textit{i.e.} of varying resistance of the individual components of the network) could also prove useful for sensor applications, to localise the most active sensing areas in the network.\\

\newpage
\begin{appendices}
\appendix
\section*{Appendix A : Numerical calculation of the effective resistance}
\label{appendix_methods_calculation_resistance}

Several methods exist to solve Kirchhoff laws, written in a matrix form, $\vec{\hat{I}}=\hat{\mathcal{L}} \vec{\hat{V}}$. Direct solver, or iterative solvers (when the dimension of the conductance matrix increases) as the preconditionned conjugate gradient method are well-known methods to solve linear systems and have already been used to compute the effective resistance of random percolating networks \cite{Zezelj2012,Simoneau2013,Simoneau2015}. \\

The non-dimensional effective resistance can also be also obtained by a Green function approach (which amounts to invert operator $\mathcal{L}$ on the orthogonal of its kernel), also called "two-point resistance method" \cite{Wu2004}, as done by Kumar \textit{and al.} \cite{Kumar2017}. The effective resistance is thus related to the eigenvalues and to two components (corresponding to the two electrode nodes) of every associated eigenvector  of the conductance matrix :
\begin{equation}
\hat{R}^{eq} = \frac{\hat{V}}{\hat{I}}= \frac{\hat{V}_{1}-\hat{V}_{P}}{\hat{I}} = \sum_{\hat{\lambda_i} \neq 0} \frac{1}{\hat{\lambda_i}} |\mathbf{\hat{\psi}}_{i,P}-\mathbf{\hat{\psi}}_{i,1}|^2
\label{R_eq}
\end{equation}
as described by Wu in 2004 \cite{Wu2004}, or alternatively :
\begin{equation}
\hat{R}^{eq} =\left[ \mathcal{L}^{PI}\right]_{11}+\left[ \mathcal{L}^{PI}\right]_{PP}-\left[ \mathcal{L}^{PI}\right]_{1P}-\left[ \mathcal{L}^{PI}\right]_{P1} = \begin{pmatrix}
1 \\
0 \\
... \\
0 \\
-1
\end{pmatrix}^T \mathcal{L}^{PI} \begin{pmatrix}
1 \\
0 \\
... \\
0 \\
-1
\end{pmatrix} > 0
\label{R_eq_pseudo_inverse}
\end{equation}
where $\mathcal{L}^{PI}$ is the Moore-Penrose pseudo-inverse of the conductance matrix $\mathcal{L}$. Indexes $1$ and $P$ are respectively associated to the lower and higher electrodes, $\left\lbrace \hat{\lambda_i} = \hat{\lambda_i}(x,y) \right\rbrace_i$ are the eigenvalues of $\hat{\mathcal{L}}(x,y)$ and $\left\lbrace \hat{\psi_i}(x,y) \right\rbrace_i$ are a set of orthonormal associated eigenvectors, and $\hat{\psi_{i,1}}$, $\hat{\psi_{i,P}}$ their components over the two nodes representing the electrodes. \\

Here, we have used both this method \cite{Wu2004} (for small sizes of the adjacency matrix, up to a few thousand lines) and a direct linear system solver based on Cholevsky decomposition of $\mathcal{\hat{L}}$ for larger sizes (up to ten thousands lines). We asserted that the two methods yielded the same results within numerical accuracy.

\appendix
\section*{Appendix B : Non-dimensionalization, application of $\pi-$theorem}
\label{appendix_pi_theorem}

The effective resistance $R^{eq}$ of a nanowire network made up of $N$ nanowires of length $l^*$, in a square device of size $L$ depends \textit{a priori} on these three structural parameters, on top of the physical unit resistances at play (linear resistance per unit length $\rho$, wire/wire contact resistance $R_c$, electrode/ nanowire contact resistance $R_{m,w}$), so that there exists a function relating implicitly these 7 \textit{independent} variables :
\begin{equation}
    f\left( R^{eq},\rho,R_c,R_{m,w},L,l^*,N \right)=0
\end{equation}

\begin{table}[!h]\centering
\begin{tabular}{|>{\centering}m{3cm}|>{\centering}m{1cm}|>{\centering}m{1.cm}|>{\centering}m{1.cm}|>{\centering}m{1.cm}|>{\centering}m{1.cm}|>{\centering}m{1.cm}|>{\centering}m{1.cm}|}
 \hline
Fundamental unity & $R^{eq}$ & $\rho$ & $R_c$ & $R_{m,w}$ &  $L$ & $l^*$ & $N$  \tabularnewline
\hline
$M$ (kg) & 1 & 1 & 1 & 1 & 0 & 0 & 0 \tabularnewline
\hline
$L$ (m) & 2 & 1 & 2 & 2 & 1 & 1 & 0 \tabularnewline
\hline
$T$ (s) & -3 & -3 & -3 & -3 & 0 & 0 & 0 \tabularnewline
\hline
$A$ (Amperes) & -2 & -2 & -2 & -2 & 0 & 0 & 0 \tabularnewline
\hline
\end{tabular}
\caption{Decomposition of the variables at play on four fundamental physical units (a resistance is homogeneous to kg.m$^2$.s$^{-3}$.A$^{-2}$)}
\label{dimensions_variables_pi_theorem_application}
\end{table}

 The rank of the $4 \times 7$ matrix in Table \ref{dimensions_variables_pi_theorem_application} being 2, application of Buckingham $\pi$-theorem leads to $7-2=5$ non-dimensional variables, so that there exists a function $\Tilde{f}$ such that :
 \begin{equation}
     \Tilde{f}(x_1,x_2,x_3,x_4,x_5)=0
 \end{equation}
 where the $x_i$'s are independent combinations of the seven physical parameters of the form $x_i=\left( R^{eq} \right)^{a_1^i} \rho^{a_2^i} R_c^{a_3^i} R_{m,w}^{a_4^i} L^{a_5^i} (l^*)^{a_6^i} N^{a_7^i}$ where the exponents $(a_k^i)_{k=1..7}$ are integers.\\
 
 \begin{table}[!h]\centering
\begin{tabular}{|>{\centering}m{3cm}|>{\centering}m{1cm}|>{\centering}m{1.cm}|>{\centering}m{1.cm}|>{\centering}m{1.cm}|>{\centering}m{1.cm}|>{\centering}m{1.cm}|>{\centering}m{1.cm}|}
 \hline
Fundamental unity & $R^{eq}$ & $\rho$ & $R_c$ & $R_{m,w}$ &  $L$ & $l^*$ & $N$  \tabularnewline
\hline
$MT^{-3}A^{-2}$ (kg.s$^{-3}$.A$^{-2}$) & 1 & 1 & 1 & 1 & 0 & 0 & 0 \tabularnewline
\hline
$L$ (m) & 2 & 1 & 2 & 2 & 1 & 1 & 0 \tabularnewline
\hline
\end{tabular}
\caption{Decomposition of the variables at play on two main physical units (deduced from Table \ref{dimensions_variables_pi_theorem_application}.}
\label{dimensions_variables_pi_theorem_application_bis}
\end{table}

The construction of non-dimensional variables can be done from Table \ref{dimensions_variables_pi_theorem_application_bis}, choosing variables $\left( R^{eq} \right)^{a_1} \rho^{a_2} R_c^{a_3} R_{m,w}^{a_4} L^{a_5} (l^*)^{a_6} N^{a_7}$ where the vector $(a_1,a_2,a_3,a_4,a_5,a_6,a_7)$ is in the kernel of this $2 \times 7$ matrix. Chosing the five independent vectors $(1,0,-1,0,0,0,0)$, $(0,1,-1,0,0,1,0)$, $(0,0,-1,1,0,0,0)$ $(0,0,0,0,0,0,1)$, $(0,0,0,0,1,-1,0)$ in the kernel of this matrix gives respectively $x_1=\frac{R^{eq}}{R_c}$, $x_2=\frac{\rho l^*}{R_c}$, $x_3=\frac{R_{m,w}}{R_c}$, $x_4=N$ (which was already non-dimensional), $x_5=\frac{L}{l^*}$ \textit{i.e.} :
\begin{equation}
     \Tilde{f} \left(\frac{R^{eq}}{R_c},\frac{\rho l^*}{R_c},\frac{R_{m,w}}{R_c},N,\frac{L}{l^*} \right)=0
 \end{equation}

\appendix
\section*{Appendix C : $\hat{R}^{eq}(x,y)$ is a rational fraction in $x$ and $y$ in the most general case}
\label{R_eq_x_y_rational_function}

Let us consider the slightly perturbed system of non-dimensional Kirchoff's laws $\vec{\hat{I}}=\left(\hat{\mathcal{L}}(x,y) +\epsilon I_P \right) \vec{\hat{V^{\epsilon}}}$, where $I_P$ is the identity matrix, $\epsilon$ is small enough and varies on a range $]0,\alpha[$ such that $\det\left(\hat{\mathcal{L}}(x,y) +\epsilon I_P \right) \neq 0$. From Cramer's formula, the voltages $V_1^{\epsilon}(x,y)$ and $V_P^{\epsilon}(x,y)$ solution of this system can be written as $\frac{\det\left( \hat{\mathcal{L}}_k(x,y)+\epsilon(I_P-\delta_{kk}) \right)}{\det\left( \hat{\mathcal{L}}(x,y) +\epsilon I_P \right)}$, with $k=1$ (respectively $k=P$), $\hat{\mathcal{L}}_k(x,y)$ being the matrix $\hat{\mathcal{L}}$ with column $k$ replaced by the vector $\vec{\hat{I}} = (\frac{I}{I_0},0,..,0,-\frac{I}{I_0})$, and $\delta_{kk}$ the matrix with $1$ at the $(k,k)$ entry and $0$ elsewhere. Given that all coefficients of $\hat{\mathcal{L}}(x,y)$ are rational functions of $x$ and $y$ from equation \ref{M*_dependances_bis2}, $\hat{V_1^{\epsilon}}(x,y)$ and $\hat{V_P^{\epsilon}}(x,y)$ are rational functions of $\epsilon$, $x$ and $y$, yielding a rational function of $x$ and $y$ when $\epsilon \rightarrow 0$, in particular for the effective resistance $\hat{R}^{eq}(x,y)=\lim_{\epsilon \rightarrow 0} \left(\hat{V_1^{\epsilon}}(x,y)-\hat{V_P^{\epsilon}}(x,y) \right)$ (taking $I=I_0$). \\
Let us point out that Kagan already noticed that the effective resistance of a resistor network could be derived in a closed-form applying Cramer's formula to the linear system \cite{Kagan2015}.

\end{appendices}

\newpage

\begin{thebibliography}{10}

\bibitem{Rocha2015}
Claudia~Gomes da~Rocha, Hugh~G. Manning, Colin O{\textquotesingle}Callaghan,
  Carlos Ritter, Allen~T. Bellew, John~J. Boland, and Mauro~S. Ferreira.
\newblock Ultimate conductivity performance in metallic nanowire networks.
\newblock {\em Nanoscale}, 7(30):13011--13016, 2015.

\bibitem{Li2007}
Jiantong Li, Zhi-Bin Zhang, and Shi-Li Zhang.
\newblock Percolation in random networks of heterogeneous nanotubes.
\newblock {\em Applied Physics Letters}, 91(25):253127, dec 2007.

\bibitem{Li2008}
Jiantong Li, Zhi-Bin Zhang, Mikael \"Ostling, and Shi-Li Zhang.
\newblock Improved electrical performance of carbon nanotube thin film
  transistors by utilizing composite networks.
\newblock {\em Applied Physics Letters}, 92(13):133103, mar 2008.

\bibitem{Li2009}
Jiantong Li and Shi-Li Zhang.
\newblock Understanding doping effects in biosensing using carbon nanotube
  network field-effect transistors.
\newblock {\em Physical Review B}, 79(15), apr 2009.

\bibitem{OCallaghan2016}
Colin O{\textquotesingle}Callaghan, Claudia~Gomes da~Rocha, Hugh~G. Manning,
  John~J. Boland, and Mauro~S. Ferreira.
\newblock Effective medium theory for the conductivity of disordered metallic
  nanowire networks.
\newblock {\em Physical Chemistry Chemical Physics}, 18(39):27564--27571, 2016.

\bibitem{Hecht2006}
David Hecht, Liangbing Hu, and George Gr\"uner.
\newblock Conductivity scaling with bundle length and diameter in single walled
  carbon nanotube networks.
\newblock {\em Applied Physics Letters}, 89(13):133112, sep 2006.

\bibitem{Kumar2017a}
Ankush Kumar.
\newblock Predicting efficiency of solar cells based on transparent conducting
  electrodes.
\newblock {\em Journal of Applied Physics}, 121(1):014502, jan 2017.

\bibitem{Nirmalraj2009}
Peter~N. Nirmalraj, Philip~E. Lyons, Sukanta De, Jonathan~N. Coleman, and
  John~J. Boland.
\newblock Electrical connectivity in single-walled carbon nanotube networks.
\newblock {\em Nano Letters}, 9(11):3890--3895, nov 2009.

\bibitem{Forro2018}
Csaba Forr{\'{o}}, L{\'{a}}szl{\'{o}} Demk{\'{o}}, Serge Weydert, J{\'{a}}nos
  Vörös, and Klas Tybrandt.
\newblock Predictive model for the electrical transport within nanowire
  networks.
\newblock {\em {ACS} Nano}, 12(11):11080--11087, nov 2018.

\bibitem{Kocabas2007}
C.~Kocabas, N.~Pimparkar, O.~Yesilyurt, S.~J. Kang, M.~A. Alam, and J.~A.
  Rogers.
\newblock Experimental and theoretical studies of transport through large
  scale, partially aligned arrays of single-walled carbon nanotubes in thin
  film type transistors.
\newblock {\em Nano Letters}, 7(5):1195--1202, may 2007.

\bibitem{Simoneau2013}
Louis-Philippe Simoneau, J{\'{e}}r{\'{e}}mie Villeneuve, Carla~M. Aguirre,
  Richard Martel, Patrick Desjardins, and Alain Rochefort.
\newblock Influence of statistical distributions on the electrical properties
  of disordered and aligned carbon nanotube networks.
\newblock {\em Journal of Applied Physics}, 114(11):114312, sep 2013.

\bibitem{Simoneau2015}
Louis-Philippe Simoneau, J{\'{e}}r{\'{e}}mie Villeneuve, and Alain Rochefort.
\newblock Electron percolation in realistic models of carbon nanotube networks.
\newblock {\em Journal of Applied Physics}, 118(12):124309, sep 2015.

\bibitem{Lyons2008}
Philip~E. Lyons, Sukanta De, Fiona Blighe, Valeria Nicolosi, Luiz Felipe~C.
  Pereira, Mauro~S. Ferreira, and Jonathan~N. Coleman.
\newblock The relationship between network morphology and conductivity in
  nanotube films.
\newblock {\em Journal of Applied Physics}, 104(4):044302, aug 2008.

\bibitem{Mutiso2013}
Rose~M. Mutiso and Karen~I. Winey.
\newblock Electrical percolation in quasi-two-dimensional metal nanowire
  networks for transparent conductors.
\newblock {\em Physical Review E}, 88(3), sep 2013.

\bibitem{Behnam2007}
Ashkan Behnam and Ant Ural.
\newblock Computational study of geometry-dependent resistivity scaling in
  single-walled carbon nanotube films.
\newblock {\em Physical Review B}, 75(12), mar 2007.

\bibitem{Zezelj2012}
Milan {\v{Z}}e{\v{z}}elj and Igor Stankovi{\'{c}}.
\newblock From percolating to dense random stick networks: Conductivity model
  investigation.
\newblock {\em Physical Review B}, 86(13), oct 2012.

\bibitem{Wu2004}
F~Y Wu.
\newblock Theory of resistor networks: the two-point resistance.
\newblock {\em Journal of Physics A: Mathematical and General},
  37(26):6653--6673, jun 2004.

\bibitem{Kagan2015}
Mikhail Kagan.
\newblock On equivalent resistance of electrical circuits.
\newblock {\em American Journal of Physics}, 83(1):53--63, jan 2015.

\bibitem{Kim2018}
Dongjae Kim and Jaewook Nam.
\newblock Systematic analysis for electrical conductivity of network of
  conducting rods by kirchhoff{\textquotesingle}s laws and block matrices.
\newblock {\em Journal of Applied Physics}, 124(21):215104, dec 2018.

\bibitem{Kirkpatrick1973}
Scott Kirkpatrick.
\newblock Percolation and conduction.
\newblock {\em Reviews of Modern Physics}, 45(4):574--588, oct 1973.

\bibitem{Heitz2011}
J{\'{e}}r{\^{o}}me Heitz, Yann Leroy, Luc H{\'{e}}brard, and Christophe
  Lallement.
\newblock Theoretical characterization of the topology of connected carbon
  nanotubes in random networks.
\newblock {\em Nanotechnology}, 22(34):345703, jul 2011.

\bibitem{Li2009a}
Jiantong Li and Shi-Li Zhang.
\newblock Finite-size scaling in stick percolation.
\newblock {\em Physical Review E}, 80(4), oct 2009.

\bibitem{Korniss2006}
G.~Korniss, M.B. Hastings, K.E. Bassler, M.J. Berryman, B.~Kozma, and
  D.~Abbott.
\newblock Scaling in small-world resistor networks.
\newblock {\em Physics Letters A}, 350(5-6):324--330, feb 2006.

\bibitem{Michelis2015}
Fulvio Michelis, Laurence Bodelot, Yvan Bonnassieux, and B{\'{e}}reng{\`{e}}re
  Lebental.
\newblock Highly reproducible, hysteresis-free, flexible strain sensors by
  inkjet printing of carbon nanotubes.
\newblock {\em Carbon}, 95:1020--1026, dec 2015.

\bibitem{Kumar2017}
Ankush Kumar, N.~S. Vidhyadhiraja, and Giridhar~U. Kulkarni.
\newblock Current distribution in conducting nanowire networks.
\newblock {\em Journal of Applied Physics}, 122(4):045101, jul 2017.

\end{thebibliography}

\end{document}